\documentstyle[12pt]{article}
\begin{document}
\baselineskip=8 mm plus 1 mm minus 1 mm

\begin{center}
{\bf FAMILON MODEL OF DARK MATTER}
\vspace*{.4cm}

{V.BURDYUZHA ${}^{a*}$, O.LALAKULICH ${}^{b}$, Yu.PONOMAREV ${}^{a}$, G.VERESHKOV ${}^{b}$}

\vspace*{.4cm}

\noindent
${}^{a}$ Astro Space Center of Lebedev Physical Institute, Russian Academy
of Sciences, Profsoyuznaya str.84/32, 117997 Moscow, Russia
${}^{b}$ Institute of Physics, Rostov State University, Stachki str.194,
344104 Rostov/Don, Russia\\

\end{center}

\baselineskip=4 mm plus 1 mm minus 1 mm

\vspace*{0.4cm}

If the next fundamental level of matter occurs (preons) then dark matter must
consist of familons containing a "hot" component
from massless particles and a "cold" component from massive particles. During
evolution of the Universe this dark matter was undergone to late-time
relativistic phase transitions temperatures of which were different.
Fluctuations created by these phase transitions have had a fractal character.
In the result of the structurization  of dark matter (and therefore the
baryon subsystem) has taken place and in the Universe some characteristic
scales which have printed this phenomenon arise naturally. Familons are
collective excitations of nonperturbative preon condensates which could be
produced during more early relativistic phase transition. For structurization
of dark matter (and baryon component) three generations of particles are
necessary. The first generation of particles has produced the observed baryon
world. The second and third generations have produced dark matter from particles
which have appeared when symmetry among generations was spontaneously broken.

\vspace*{0.4cm}


\newpage

{\bf INTRODUCTION}

\baselineskip=8 mm plus 1 mm minus 1 mm

Using the preon structure of quarks and leptons the familon model of dark
matter (DM) proposed by Hill, Schramm and Fry(1989) and Frieman et al.(1992)
is reanimated. This model has more physical and cosmological consequences if
the next structure level of matter is involved. Our interest to the preon model of
elementary particles was also induced by the fact of possible leptoquarks
resonance in HERA experiment (Adloff et al.,1997 and Breitweg et al.1997)
the possibility to research a pair production of scalar leptoquarks at
FERMILAB TEVATRON (Kramer et al.,1997 and Affolder et al.,2001). The standard
model of physics of elementary particles is not expected to be a complete theory
(it does not explain the number of fermion families, their mass hierarchy and
does not provide a unified description of all gauge symmetries). Although, of course,
the standard model describes very well all experiments by fundamental
fermions and their interactions via gauge bosons. Compositeness models of quarks and
leptons postulate a new strong dynamics which bind constituents (preons) although
a sure motivation to these models is absent till now.
At first important cosmological and physical consequences enumerate.If DM
consist of familons then in this medium late-time phase transitions were
possible and fluctuations created by these phase-transitions have had a fractal
character. These fractal fluctuations could develop into the fractal large-
scale structure (LSS) of the baryon component. Note that the fractal structure
of the baryon component is not observed on all scales (Back and Chen,2001;
Wu et al.,1999). In the Universe these late phase transitions have also produced some
characteristic  scales. Besides, understanding of three generations of elementary
particles is become naturally. The observed baryon world and DM may
be realized only then when three generations of particles occur. The first
generation of particles has produced the baryon world. The second
and third generations have produced dark matter. The structurization of dark
matter (and the baryon component) has been produced particles appeared when the
symmetry among generations was spontaneously broken.

The preon structure of matter was introduced by Pati and Salam (1974) and
was studied by many authors (Terazawa et al.,1977; Lane et al.,1977; Terazawa,1980;
Eichten et al.,1984). We have studied the structure of preon nonperturbative
vacuum which may arise in the result of the correlation of non-Abelian fields on two
scales. $\Lambda_{mc} \gg 1 \; TeV$ is the confinement scale of metacolour
and $\Lambda_{c} \sim150 \; MeV$ is the quantum chromodynamics (QCD) scale.
We have detected that in the spectrum of excitations of heterogenic
nonperturbative preon vacuum pseudo-Goldstone modes of familon type appear.
Familons are created when the symmetry of quark-lepton generations is
spontaneously broken and their nonzero masses are the result of superweak
interactions with quark condensates. Physics of the spontaneously broken
symmetry of generations (production of familons) was considered by Feng et al.,(1998).
The distinguishing characteristic of these particles is the availability of
the residual $U(1)$ symmetry and possibility of it spontaneous breaking for
temperature $\frac{\Lambda_{mc}}{\Lambda^{2}_{c}} \sim 10^{-3} \; eV$
in the result of relativistic phase transitions.

We have proposed that these relativistic phase transitions (RPT) had
the direct relation to the production of primordial perturbations in DM
the evolution of which leads to the fractal baryon large scale structure.
Note other time that the idea of RPT in the cosmological gas of pseudo-
Goldstone bosons in connection with LSS problems was early formulated in
articles of Hill et al.,(1989) and Frieman et al.,(1992). Here we have
investigated by quantitatively the preon-familon model of these RPT.

In the beginning the astrophysical motivation of our theory is discussed
more detail. Observational data show that some baryon objects such as
the quasar on  $z \sim 4.9$ and the galaxy on $z \sim 6.68$, CO lines on
$z \sim 4.43$ and $z \sim 4.69$ (Omont et al.,1996; Chen et al.,1999) were produced as minimum on redshifts
$z \sim 6 \div 8$. This is the difficulty for the standard CDM and
$\Lambda$CDM models to produce their (the best fit is $z \sim 2 \div 3$ (Madau,1999)
and observations provide the support of this). If early baryon cosmological
structures produced on $z > 10$ then the key role must play DM particles
with nonstandard properties.

Probably DM consists of ideal gas particles with $m \approx 0$
practically noninteracting with usual matter (till now they  do not detected
because of their superweak interaction with baryons and leptons). In the
standard cosmological model DM (here CDM) contains $25\%$ of the total
density that is: $$\Omega_0 = \Omega_{\Lambda} +
\Omega_{CDM} +  \Omega_{\nu} + \Omega_{b} = 0.7 + 0.25 + 0.02 + 0.03 = 1.$$
In the article (Caldwell,2004) more exact data on $\Omega _{\Lambda}$ are given.
Also the important point is to know the end of formation of observed baryon structures.
A characteristic moment of the most of cosmological structures formation finishing
remains the same ($z \sim 2 \div 3$). The appearance of baryon structures on
high red shifts $(z > 4)$ was the result of statistical outbursts evolution of
the spectrum of DM density perturbations. Therefore early cosmological baryon
structures may be connected to statistical outbursts only in the sharply nonlinear
physical system which is a medium after RPT (the production of inhomogeneities).

Note again that we have investigated the idea in which the baryon component
of matter repeats the structure of DM owing to gravitation. That is
relativistic phase transitions have produced DM fractal fluctuations in
which baryons have subsequently clustered. The fractal structure of the baryon
component was studied by Bak and Chen (2001) and Wu et al.,(1999) starting from
the article of Coleman et al.,(1988) in which authors
have suggested that the Universe up to some Mpc has the fractal structure
(the fractal structure was observed up to 50 Mpc by Martinez and Coles (1994)
and the sharp transition to homogeneity was predicted at 300 Mpc also by these
authors. Only the critical phenomenon like to a phase transition creates fractal structures.

A new theory of DM must combine the properties of superweak interaction
of DM particles with baryons and leptons and  intensive interaction of these
particles each other. Such interactions are provided by nonlinear properties
of DM medium. This is the condition for realization of RPT.

{\bf THE BASIC ARGUMENTS}

The familon symmetry is experimentally observed (the different
generations of quarks and leptons participate in gauge interactions the
same way). Breaking of this symmetry gives masses of particles in
different generations. A hypothesis about spontaneous breaking of
familon symmetry is natural and the creation of Goldstone bosons is
inevitably. The properties of any pseudo-Goldstone bosons as and
pseudo-Goldstone bosons of familon type depend on the physical
realization of Goldstone modes. These modes may be arisen  from
fundamental Higgs fields or from collective excitations of a heterogenic
nonperturbative vacuum condensate more complex than a quark-gluon one in
QCD. The second possibility can realize the theory in which quarks and
leptons are composite particles that is the preon model of elementary
particles. If leptoquarks will be detected then two variants of
explanations may be. If a leptoquarks resonance will be narrow and high
then these leptoquarks come from GUT or SUSY theories. The low and
wide resonance can be explained by composite particles only.

The simplest boson-fermion preon model consists of left-handed fermion
preons $U^{\alpha}_{L}, D^{\alpha}_{L}$ and  scalar preons of quark
$(\Phi^{i \alpha}_{a})$ and lepton $(\chi^{\alpha}_{l})$ types. In this model
the interior structure of elementary particles is:
$$u^{i}_{La} = U^{\alpha}_{L} \Phi^{+i \alpha}_{a} \;\;\;
u^{i}_{La} = (u^{i}_{L}, c^{i}_{L}, t^{i}_{L})$$
$$d^{i}_{La} = D^{\alpha}_{L} \Phi^{+ i \alpha}_{a} \;\;\;
d^{i}_{La} = (d^{i}_{L}, s^{i}_{L}, b^{i}_{L}) \eqno(1)$$
$$\nu^{i}_{Ll} = U^{\alpha}_{L} \chi^{\alpha}_{l} \;\;\; \nu^{i}_{Ll} =
\nu_{L_{e}}, \nu_{L_{\mu}}, \nu_{L_{\tau}}$$
$$l^{i}_{Ll} = D^{\alpha}_{L} \chi^{\alpha}_{l} \;\;\; l^{i}_{Ll} =
(e_{L}, \mu_{L}, \tau_{L})$$

In the case of leptoquarks our model gives:
$$(LQ)_{al} = \Phi^{i \alpha}_{a} \chi^{\alpha}_{l} \eqno(2)$$
here and in the following $i$ is  color index of QCD, $ a, b, c=1, 2, 3;
l, m, r =1, 2, 3 $ are numbers of quark and lepton generations, $\alpha$ is
metacolor index corresponding to a new metachromodynamics interaction
linking preons in quarks and leptons.

Inside quarks and leptons metagluon fields $G^{\omega}_{\mu \nu}$ and
scalar preon fields are in the confinement state like to the confinement of
quarks and gluons inside hadrons. This effect is provided by the existence
of nonperturbative metagluon and preon condensates:
$$\langle 0 \mid \frac{\alpha_{mc}}{\pi} G^{\omega}_{\mu \nu} G^{\mu
\nu}_{\omega} \mid 0 \rangle \sim \Lambda^{4}_{mc} \eqno(3)$$
$$\langle 0 \mid \Phi^{+ i \alpha}_{a} \Phi^{i \alpha}_{b} \mid 0
\rangle = V_{ab} \sim - \Lambda^{2}_{mc} \eqno(4)$$
$$\langle 0 \mid \chi^{+ \alpha}_{l} \chi^{\alpha}_{m} \mid 0 \rangle =
V_{lm} \sim - \Lambda^{2}_{mc} \eqno(5)$$
here $\Lambda_{mc}$ is the energetic scale of preon confinement,
$V_{ab}, V_{lm}$ are the condensate matrixes. Condensates (3)
and (4) together with gluon and quark condensates ($\langle 0 \mid
\frac{\alpha_{s}}{\pi} G^{a}_{\mu \nu} G^{\mu \nu}_{a} \mid 0 \rangle;
\;\; \\
\langle 0 \mid \bar{q}_{L} q_{R} + \bar{q}_{R} q_{L} \mid 0 \rangle)$
provide the mechanism of mass quarks production of all third generations.
It is shown on the diagram 1:

\begin{picture}(350,200)
\put(50,100){\vector(1,0){10}}
\put(60,100){\line(1,0){10}}
\put(70,100){\circle*{6}}
\put(70,103){\line(1,0){35}}
\multiput(70,97)(10,0){4}{\line(1,0){5}}
\multiput(105,93)(0,-10){7}{\line(0,1){5}}
\multiput(200,150)(0,-10){13}{\line(0,1){5}}
\put(105,150){\vector(0,-1){25}}
\put(250,150){\vector(0,-1){25}}
\put(105,125){\line(0,-1){22}}
\put(250,125){\line(0,-1){25}}
\put(250,100){\vector(1,0){40}}
\put(290,100){\line(1,0){15}}
\multiput(108,115)(3,0){11}{\oval(6,3)}
\multiput(108,70)(3,0){11}{\oval(6,3)}
\multiput(150,100)(3,0){17}{\oval(6,3)}
\put(140,112) {$ \mbox{x} $}
\put(140,97) {$ \mbox{x} $}
\put(140,67) {$ \mbox{x} $}
\put(103,150) {$ \mbox{x} $}
\put(103,25) {$ \mbox{x} $}
\put(247,150) {$ \mbox{x} $}
\put(197,150) {$ \mbox{x} $}
\put(197,25) {$ \mbox{x} $}
\multiput(80,40)(3,0){8}{\oval(6,3)}
\multiput(203,40)(3,0){22}{\oval(6,3)}
\multiput(80,10)(3,0){62}{\oval(6,3)}
\multiput(80,10)(0,3){10}{\oval(3,6)}
\multiput(270,97)(0,-3){10}{\oval(3,6)}
\put(267,37) {$ \mbox{x} $}
\put(267,60) {$ \mbox{x} $}
\put(267,7) {$ \mbox{x} $}
\put(100,175) {$\langle 0 | \bar{U}^{\alpha}_{L}\phi^{r\alpha}_{c} u^r_{Rb} | 0 \rangle
\equiv
\langle 0 | \bar{u}^m_{Lc} u^m_{Rb} | 0 \rangle $}
\put(110,17) {$\langle 0 | \phi^{+k\beta}_{a} \phi^{k\beta}_{c} | 0 \rangle $}
\put(110,80) {$\langle 0 | \frac{\alpha_{mc}}{\pi}
G^{\alpha \gamma}_{\mu \rho} G^{\gamma \beta}_{\rho \nu}
G^{\beta \alpha }_{\nu \mu} | 0 \rangle $}
\put(255,50) {$\langle 0 | \frac{\alpha_{s}}{\pi}
G^{ik}_{\mu \nu} G^{kr}_{\nu \rho}
G^{ri}_{\rho \mu} | 0 \rangle $}
\put(50,110) {$\bar{u}^{i}_{La} $}
\put(280,110) {$u^{i}_{Rb} $}
\put(80,110) {$\bar{U}^{\gamma}_{L} $}
\put(80,80) {$\phi^{i\gamma}_{a} $}
\put(170,-10){$ \mbox{Diagram 1} $}
\end{picture}
\vspace*{.5cm}

\noindent
in which  $ G^{ik}_{\mu \nu} = \lambda^{ik}_{n} G^{in}_{\mu \nu}, \;\;
\lambda^{ik}_{n}$ is Gell-Mann matrices;
$G^{\alpha \beta}_{\mu \nu} =
\lambda^{\alpha \beta}_{\omega} G^{\omega}_{\mu \nu}, \;
\lambda^{\alpha \beta}_{\omega}$ is an analogue of Gell-Mann matrices for metacolour.
As it can see from this diagram (1) the main contribution in the effect
of familon symmetry vacuum breaking is formed by the preon condensates (4).

The theory of preons predicts the complex structure of a heterogenic
nonperturbative vacuum and familons are collective excitations of these
condensates. These excitations are the result of local processes of
weakening and rebuilding of correlations among fields entering in condensates:
$$M^{(u)}_{ab} = \langle 0 \mid \Phi^{\alpha k}_{a} \Phi^{\alpha k}_{c}
\bar{U}^{\beta}_{L} \Phi^{\beta_{i}}_{c} q^{i}_{R_{b}} \mid 0 \rangle \eqno(6)$$
$$M^{(d)}_{ab} = \langle 0 \mid \Phi^{\alpha k}_{a} \Phi^{\alpha k}_{c}
\bar{D}^{\beta}_{L} \Phi^{\beta_{i}}_{c} q^{i}_{R_{b}} \mid 0 \rangle \eqno(7)$$
$$M^{(l)}_{lm} = \langle 0 \mid \chi^{\alpha}_{l} \chi^{\alpha}_{r}
\bar{D}^{\beta}_{L} \chi^{\beta}_{r} l_{Rm} \mid 0 \rangle \eqno(8)$$
Also it is necessary to note the peculiar properties of the first generation of
quarks. Their masses are exclusively produced by the interaction with the
quark-gluon condensate. The production of second and third generations of quark masses
is outside limits of QCD. But this fact may be natural in preon model. Scalar
preon condensates of the first generation are efficiently suppressed and they
do not carry contribution in (6-8). This situation may be explained in the
model containing composite scalar preons (the scale more than $\Lambda_{mc}$).
In this preon -subpreon model (Evnin,1997) the initial familon symmetry
$SU_{F}(3) \rightarrow SU_{F}(2)$ is broken on scale $\Lambda_{smc} \gg \Lambda_{mc}$
and then on more low scale the symmetry $SU_{F}(2) \rightarrow U(1)$ is broken
also. Therefore here we will discuss the chiral-familon symmetry of second and
third generations only ( the discussion of the familon symmetry was in detail given by
Feng (1998). 3 types of nonperturbative condensates correspond to 3 type of familon
fields and a number of familons of every type equals 8. In each type 2 familon fields
arise as the local perturbation of a condensate energy density. The rest 6 familon
fields arise as the result of a condensate rebuilding.

Thus, in the frame of preon theory DM is interpreted as the system of
familon collective excitations of the heterogenic nonperturbative vacuum.
This system consists of 3 subsystems:

1) familons of up-quark type;

2) familons of down-quark type;

3) familons of lepton type.

On stages of the cosmological evolution when $T \ll \Lambda_{mc}$ the heavy
unstable familons are absent. Small masses of familons are the result
of superweak interactions of Goldstone fields with nonperturbative vacuum
condensates and therefore familons acquire status of pseudo-Goldstone bosons.
The value of these masses is limited by the astrophysical and laboratory
magnitudes (Groom,2000).

$$ \left\{
\begin{array}{ r c l}
m_{astrophysical} & \sim & 10^{-3} \div 10^{-5} \; eV \\
m_{laboratory}    & \le  & 10 \; eV  \\
\end{array}
\right.  \eqno(9)
$$

The effect of familons mass production corresponds formally mathematically the
appearance of mass terms in the Lagrangian of Goldstone fields. From general
considerations one can propose that massive terms may arise as with "right"
as and with "wrong" signs. The sign of the massive terms predetermines the destiny
of residual symmetry of Goldstone fields. In the case of "wrong" sign for low
temperatures $T <T_{c} \sim m_{familons} \sim 0.1 \div  10^{5} \; K$ a Goldstone
condensate produces and the symmetry of familon gas breaks
spontaneously.

The representation about physical nature of familon  excitations described
above is formalized in a theoretical-field model. As example we discuss the model
only one familon subsystem corresponding to up-quarks of second and third generations.
The chiral-familon group of the model is $SU_{L}(2) \times SU_{R}(2)$.
The familon excitations are described by an eight measure (on number of matrix
components (6)) reducible representation of this group factorized on two irreducible
representations $(F, f_{a}); \; (\psi, \varphi_{a})$ which differ each other by
a sign of space chirality. In this model the interaction of quark fields with
familons occurs. However in all calculations quark fields are represented in the
form of nonperturbative quark condensates. From QCD and the experiment the
connection between quark and gluon condensates is known:

$$\langle 0 \mid \bar{q} q \mid 0 \rangle \approx \frac{1}{12 m_{q}}
\;\;\;\;\; \langle 0 \mid \frac{\alpha_{s}}{\pi} G^{n}_{\mu \nu} G^{\mu
\nu}_{n} \mid 0 \rangle \approx \frac{3 \Lambda^{4}_{c}}{4 m_{q}} \eqno(10)$$
here: $q = t, c; \;\; m_{c} \sim 1.5 \; Gev; \;\; m_{t} \sim 175 \; Gev; \;\;
\Lambda_{c} \sim 150 \; Mev$.

The spontaneous breaking of symmetry $SU_{L}(2) \times SU_{R}(2)
\rightarrow U(1)$ is produced by vacuum shifts $\langle \psi \rangle =
v; \; \langle f_{3} \rangle = u$. The numerical values $v, u \sim
\Lambda_{mc}$ are unknown. They can be found by experimentally if
our theory corresponds to reality. Parameters $\; u$ and $v$ together with
the value of condensates (10) define numerical values of basic
magnitudes characterizing the familon subsystem. After
breaking of symmetry  $SU_{L}(2) \times SU_{R}(2) \rightarrow U(1)$
light pseudo-Goldstone fields contain the real pseudoscalar field with
the mass:
$$m^{2}_{\varphi^{'}} = \frac{1}{6(u^{2} + v^{2})} \langle 0 \mid
\frac{\alpha_{s}}{\pi} G^{n}_{\mu \nu} G^{\mu \nu}_{n} \mid 0 \rangle \eqno(11)$$
the complex pseudoscalar field with the mass:
$$m^{2}_{\varphi} = \frac{1}{24 v^{2}} \frac{m_{t}}{m_{c}} \langle 0
\mid \frac{\alpha_{s}}{\pi} G^{n}_{\mu \nu} G^{\mu \nu}_{n} \mid 0
\rangle \eqno(12)$$
and the complex scalar field the mass square of which is negative:
$$m^{2}_{f} = - \frac{1}{24 u^{2}} \frac{m_{t}}{m_{c}} \langle 0 \mid
\frac{\alpha_{s}}{\pi} G^{n}_{\mu \nu} G ^{\mu \nu}_{n} \mid 0 \rangle \eqno(13)$$

The complex field with masses (12-13) is the nontrivial representation of
residual symmetry of $U(1)$ group but the real field (11) is the sole
representation of this group. We propose that cosmological DM consists
of particles with these masses and their analogies from the down-
quark-familons and the lepton-familon subsystems.

The negative mass square of complex scalar field means that for
$$T < T_{c(up)} \sim \mid \bar{m}_{f} \mid \sim
\frac{\Lambda_{mc}}{\Lambda^{2}_{c}} \sqrt{m_{t}/m_{c}} \eqno(14)$$
pseudo-Goldstone vacuum is unstable that is when $T = T_{c(up)}$ in gas
of pseudo-Goldstone bosons should be  RPT in the state with spontaneous
breaking $U(1)$ symmetry. Two other familon subsystems can be studied by the
same methods. Therefore DM consisting of pseudo-Goldstone bosons of familon type
is a many component heterogenic system evolving complex thermodynamical way.

In the phase of breaking symmetry every complex field with masses (12-13)
splits on two real fields with different masses. That is the familon
subsystem of up-quark type consists of five kinds of particles with
different masses. Analogous phenomenon takes place in the down-quark
subsystem. The breaking of residual symmetry is when:
$$T_{c(down)} \sim \frac{\Lambda_{mc}}{\Lambda^{2}_{c}}
\sqrt{m_{b}/m_{s}} \eqno(15)$$
In a low symmetric phase this subsystem consists also of five kinds
particles with different masses. In our theory the lepton-familon subsystem
can be undergone to RPT also but leptonic condensates are elements of new physics
which may come in the future and probably their discussion is prematurely.

{\bf RESULTS}

The relativistic phase transitions in familon subsystems must be
described in the frame of temperature quantum field theory. It is important
to underline that sufficiently strong interactions of familons each
other provide the evolution of familon subsystem through state of local
equilibrium type. Our estimates have shown that the transition in
nonthermodynamical regime of evolution occurs on stage after RPT even
if RPT took place for temperature $\sim 10^{-3} \; eV$. The thermodynamics of
a familon system may be formulated in the approximation of a self-coordinated
field. The methods of  RPT theory which were be used by us are similar to ones
of our article (Vereshkov and Burdyuzha,1995). The  nonequilibrium Landau
functional of states $F(T, \eta, m_{A})$ depends on  the order parameter
$\eta$ and five effective masses of particles $m_{A}, \; A = 1, 2, 3, 4, 5$:

$$F(T, \eta, m_{A}) = - \frac{1}{3} \sum_{A} J_{2} (T, m_{A}) + U(\eta,
m_{A}) \eqno(16)$$

here $J_{2}$ is the characteristic integrals (similar integrals
used for the description of RPT in article (Burdyuzha et al.,1997).
The conditions of the extremum of this functional on effective masses give
the equation of connection $m_{a} = m_{a}(\eta, T)$ which  defines formally
the typical functional Landau $F(T, \eta)$. The condition of minimum of
this functional on the parameter of order ${\eta}$:

$$\frac{d^{2}F}{d \eta^{2}} = \frac{\partial^{2}F}{\partial \eta^{2}} +
\sum_{A} \frac{\partial^{2}F}{\partial \eta \partial m_{A}}
(\frac{\partial m_{A}}{\partial \eta}) > 0 \eqno(17)$$

is concordant with the equation of state $\partial F/ \partial \eta =
0$ that allows: a) to establish  the kind of RPT, b) to find the
thermodynamical  boundary of stability phases, c) to calculate values
of observed magnitudes (energy density, pressure, thermal capacity,
sound velocity et al.) in each phase. More detail the thermodynamics
of the familon system has been discussed in article (Burdyuzha et al.,1998).

We have detected that RPT in familon gas is one of the first kind with wide
region of phases coexistence. Therefore in epoch of RPT or more
exactly in the region of phases coexistence the Universe had a block-phase
structure containing domains of different phases. The numerical
modelling of this RPT has shown that average contrast of density in the
block-phase structure is $\delta \epsilon/  \epsilon \sim 0.1$. This
structure is illustrated on diagram 2 in some conditional dimensionless
units.

\begin{center}
\begin{picture}(320,225)
\vspace{12cm}
\put(20,30){\line(1,0){300}}
\put(20,30){\line(0,1){175}}
\put(103,180){$T_{c(1)}$}
\put(233,180){$T_{c(2)}$}
\put(140,100){$HS - phase$}
\put(40,60){$LS - phase$}
\put(20,30){\line(3,1){220}}
\put(110,75){\line(4,1){200}}
\multiput(20,30)(0,50){4}{\line(-1,0){5}}
\multiput(20,30)(50,0){7}{\line(0,-1){5}}
\multiput(110,30)(0,10){15}{\line(0,1){5}}
\multiput(240,30)(0,10){15}{\line(0,1){5}}
\put(0,30){$0.0$}
\put(0,80){$0.1$}
\put(0,130){$0.2$}
\put(0,180){$0.3$}
\put(10,16){$0.00$}
\put(60,16){$0.01$}
\put(110,16){$0.02$}
\put(160,16){$0.03$}
\put(210,16){$0.04$}
\put(260,16){$0.05$}
\put(310,16){$0.06$}
\put(325,32){$P$}
\put(10,210){$\varepsilon (P)$}
\put(155,-5){$ \mbox{Diagram 2}$}
\end{picture}
\end{center}

\noindent
The size of domains and masses of baryon and dark matter inside
domains are defined by distance to horizon of events $L_{horiz.}$ at the
moment of RPT. As it is seen from (14-15) numerical values of these
magnitudes which are important for LSS theory depend on a value of the
unknown today parameter of the preon confinement $\Lambda_{mc}$.

If inhomogeneities appearing during RPT in familon gas have the relation to
observable scales of LSS (10 Mpc) then $\Lambda_{mc} \sim 10^{5} \; TeV$. More
detail estimates today is premature but it is necessary to note that
suggested theory contains as minimum two phase transitions and therefore two
characteristic scales of baryon LSS. Now it would be a speculation to define
exactly magnitude of these scales (probably galaxies and clusters of galaxies)
since we do not know familon masses.
Numerical estimates of inhomogeneities parameters arising as the result of
strong interaction of domains LS and HS phases in the region of their contact
show that the density contrast may increase to $\delta \epsilon/ \epsilon
\sim 1$ on the scale $L \sim 0.1 L_{horison}$ at the moment of the phase
transition and besides effects connected with fragmentation of DM medium
may be superimposed at the spectrum of the CMB radiation.

Note that Hill, Schramm, and Fry (1989) have even proposed some laboratory tests for
verification of the late-time phase transitions model (the neutrino-schison
model). Their model as and our one can potentially generate structures
(baryon and DM) at red shifts $ z>10 $.  Besides, if the fractal structure of
the baryon component will be proved finally then the late-time phase
transitions model becomes automatically the main one for the production of the
baryon LSS. Since only phase transitions realize a fractal structure for seeds.
Probably during evolution of baryon structures their fractal distribution is
smoothed down and it is not observed on large scales although there is general
agreement about the existence of fractal galactic structures at moderate
scales (Bak and Chen(2001); Wu et al.,(1999); Guzzo(1991)).

Finally note,that for structurization of DM (and baryon component of the Universe)
three generations of particles are necessary obligatory. The first generation of
particles has produced the baryon world which is observed. The second and third
generations of particles have produced a fractal distributed DM from familons
(the baryon component has repeated this distribution). Our first conclusion is that
the preon structure (the next structural level of matter) must be detected since
only the preon model (more exactly phase transitions) may provide a fractal
distribution as DM as and baryon component. Only in the preon model some scales
could be naturally produced during evolution of the Universe. Of course, the familon DM is difficult to detect owing to a
superweak interaction with usual matter. Recent search of familons by CLEO collaboration
gave the negative result (Ammar et al.,2001). Last publications on research of
DM can be found in some last reports (Caldwell,2004; Adashe and Servant,2004;
Chen,2004; Ellis,(2005); Kunz,(2007); Moffat,2004; Sahni,2004) and on site:http://www.phys.ufl.edu/~axion/welcome.html

{\bf REFERENCES}\\
Adashe K. and Servant G.(2004) hep-ph/0403143.\\
Adloff C. and co-authors (HERA collaboration)(1997) Z. Phys. C74, 191.\\
Affolder T. and co-authors (FERMILAB collaboration)(2001) Phys. Rev. Lett.87, 231803.\\
Ammar R. and co-authors (CLEO collaboration)(2001) Phys. Rev. Lett. 87, 271801.\\
Bak P. and Chen K. (2001) Phys. Rev. Lett. 86, 4215.\\
Breitweg J. and co-authors (HERA collaboration)(1997) Z. Phys. C74, 207.\\
Burdyuzha V.V., Lalakulich O.D., Ponomarev Yu.N. and Vereshkov G.M.\\
(1997) Phys. Rev. D 55, 7340 R; (1997) Astronomy Reports 42, 711.\\
Burdyuzha V.V., Lalakulich O.D., Ponomarev Yu.N. and Vereshkov G.M.\\
\hspace*{0.2cm} (1998) Preprint of Yukawa Inst. for Theoretical Physics (YITP-98-51); hep-ph/9907531.\\
Caldwell R. (2004) Phys. World 5, 37.\\
Chen H-W., Lanzetta K.M. and Pascarelle S. (1999) Nature 398, 586.\\
Chen R. and co-authors (2004) astro-ph/0403352.\\
Coleman P.H., Pietronero L. and Sanders R.H. (1988) Astron. Astrophys. 200, L32.\\
Eichten E. and co-authors (1984) Rev. Mod. Phys. 56, 579.\\
Ellis J.and co-authors (2005) hep-ph/0502001.\\
Evnin O.E. (1997) hep-ph/9711433.\\
Feng J.L. and co-authors (1998) Phys. Rev. D57, 5875.\\
Frieman J.A., Hill C.T. and Watkins R. (1992) Phys. Rev. D46, 1226.\\
Groom D.E. and co-authors (Particle Data Group) (2000) Eur. Phys. J. C15, 1;\\
http://pdg.lbl.gov (page 15).\\
Guzzo L. and co-authors (1991) Astrophys. J. Lett. 382, L5.\\
Hill C.T., Schramm D.N. and Fry J.N. (1989) Comments of Nucl.and Part. Phys. 19, 25.\\
Kramer M. and co-authors (FERMILAB collaboration)(1997) Phys. Rev. Lett. 79, 341.\\
Kunz M. (2007) arXiv:/07105712.\\
Lane K.D. and co-authors (1977) Phys. Rep. 278, 291.\\
Madau P. (1999) in Proceeding of Nobel Symposium Particle Physics.\\
Preprint N 1333 of Space Telescope Science Institute.\\
Martinez V.J. and Coles P. (1994) Astrophys. J. 437, 550.\\
Moffat J.W. (2004) hep-ph/0403266.\\
Omont A. and co-authors (1996) Nature 382, 428.\\
Pati J.C. and Salam A. (1974) Phys. Rev. D10, 275.\\
Sahni V. (2004) astro-ph/0403324.\\
Terazawa H., Chikashige Y. and Akama K. (1977) Phys. Rev. D15, 480.\\
Terazawa H. (1980) Phys. Rev. D22, 184.\\
Vereshkov G.M. and Burdyuzha V.V. (1995) Int. J. Modern. Phys. A10, 1343.\\
Wu K.K.S., Lahav O. and Rees M.J. (1999) Nature 397, 225.\\

\end{document}